\documentclass[conference]{IEEEtran}
\ifCLASSINFOpdf
\else
\fi

\usepackage[square, comma, sort&compress,numbers]{natbib}
\usepackage{flushend}
\usepackage{graphicx}

\usepackage[linesnumbered,lined,ruled,commentsnumbered]{algorithm2e}
\usepackage{amsmath}

\usepackage{setspace}
\begin{document}
\bstctlcite{IEEEexample:BSTcontrol}

\title{ Energy Saving Techniques for Phase Change Memory (PCM)}

 \author{\IEEEauthorblockN{Sparsh Mittal }
 \IEEEauthorblockA{Department of Electrical and Computer Engineering \\
 Iowa State University, Ames, Iowa 50011, USA\\
  Email: sparsh0mittal@gmail.com}
 }


\maketitle

\begin{abstract}
In recent years, the energy consumption of computing systems has increased and a large fraction of this energy is consumed in main memory. Towards this, researchers have proposed use of non-volatile memory, such as phase change memory (PCM), which has low read latency and power; and nearly zero leakage power. However, the write latency and power of PCM are very high and this, along with limited write endurance of PCM present significant challenges in enabling wide-spread adoption of PCM. To address this, several architecture-level techniques have been proposed.  In this report, we review several techniques to manage power consumption of PCM. We also classify these techniques based on their characteristics to provide insights into them. The aim of this work is encourage researchers to propose even better techniques for improving energy efficiency of PCM based main memory.  
\end{abstract}


\begin{IEEEkeywords}
Phase change memory (PCM), write power, power management, energy saving, architectural techniques.

\end{IEEEkeywords}

%
\IEEEpeerreviewmaketitle



\section{Introduction}
Modern computing systems provide several 24$\times$7 services and execute data-centric computing applications \cite{pande2013efficient,lee2011twitter,pande2011quality123}, which require large amount of computing resources. Since disk-access is slow, the amount of main memory used in computing systems has increased. However, due to this, the contribution of main memory in system power consumption has also increased, which could be as large as 40\% \cite{lefurgy2003energy}.  Also, future scaling of DRAM  is in doubt \cite{hay2011preventing} and it low density obstructs design of large sized cache using DRAM.

To address this, researchers have explored alternative memory design technologies such as phase change memory (PCM) \cite{qureshi2011phase} etc., which provide high storage density. PCM is a non-volatile device which has high retention property (over 10 years) and very good operation characteristics and scalability \cite{raoux2008phase}. PCM prototypes with feature size as small as 3nm have been fabricated \cite{raoux2008phase}. It also has higher write endurance than the flash memory \cite{cho2009flip,dhiman2009pdram}, although its write endurance is several orders of magnitude worse than that of DRAM. The read power and delay of PCM are in the same range as that of DRAM, however, its write power is significantly higher than that of DRAM  \cite{yang2007low,zhou2009durable}. Thus,  power management of PCM is extremely important to ensure its wide-spread adoption and also avoid the problems due to heating of the system.   

In this report, we review several architectural techniques for managing power consumption of PCM.  We classify these techniques based on their properties to highlight their similarities and differences. The aim of this work is to provide a synthetic overview of the research field of PCM power management. We believe that this overview will encourage researchers to propose even better techniques for improving energy efficiency of PCM based main memory.

The rest of the report is structured as follows. Section \ref{sec:background} provides a background on power management in PCM devices. Section \ref{sec:est} discusses some energy saving techniques in detail. We only discuss the essential idea of the techniques and not the quantitative results, since different techniques have been evaluated using different platforms. Finally, Section \ref{sec:conclusion} discusses the future work and provides the conclusion. 


\section{Background and Related Work}\label{sec:background}
\subsection{Need of Power Management in Main Memory}
 In recent years, the total power consumption of embedded systems, data centers and supercomputers has significantly increased \cite{Mit_DRAMsurvey,pande2012embedded}, which has also increased the carbon footprint of IT. To address this, computer architects have proposed several techniques for reducing the energy consumption of computing systems \cite{mittal2013PhDThesis,MitZha12_EnCache}.  Since a large fraction (upto 40\% \cite{dhiman2009pdram,Mit_DRAMsurvey,barroso2009datacenter}) of energy spent in server-class systems is consumed by the main memory, the techniques for saving energy in main memory systems are vital for improving energy efficiency of computing systems. 
 
 
 Since PCM has much smaller leakage energy than DRAM, recently, there has been a significant interest in use of PCM. PCM has been evaluated in context of GPUs (graphics processing unit) \cite{wang2013can,kim2012hybrid}, embedded systems \cite{shao2012utilizing,Hu2013WAR,6513577,6164966,hu2012optimizing}, real-time systems \cite{zhou2011real},   video applications \cite{fang2012softpcm,kwon2012optimizing} and so on. In fact, use of PCM has also been explored for designing caches \cite{mangalagiri2008low,joo2010energy,li2011exploiting,wang2013i2wap,guo2012wear}. As the use of PCM increases, managing its power consumption becomes much more important.

\subsection{A Brief Background on Phase Change Memory}
We briefly review the design of PCM. A PCM cell comprises an NMOS access transistor and a storage resistor which is made of a chalcogenide alloy \cite{tominaga1997structure,ramos2011page}. To store a binary value  on PCM, heat is applied to it which transitions the physical state of the alloy with particular resistances. When the alloy is heated to a very high temperature (greater than 600 degree Celsius) and quickly cooled down, it transitions into an amorphous substance with high electrical resistance which represents binary ``0''. On the other hand, when the alloy is heated to a temperature between the crystallization (300 degree Celsius) and melting (600 degree Celsius) points and cools down slowly, it crystallizes to a physical state with lower resistance, which represents binary ``1''. The difference in resistance values between the two states of PCM is typically 3 orders of magnitude. PCM memories achieve high density by  exploiting this high resistance range to store multiple bits in a single cell, this structure is known as multi-level cell or MLC. PCM is byte-addressable and is immune to radiation-induced soft errors.

\subsection{An Overview of Energy Saving Techniques for PCM}
Computer architects have proposed several techniques for saving energy in PCM systems. Some researchers have proposed hybrid PCM-DRAM design \cite{park2011power,wang2013can,zhang2009exploring,qureshi2009scalable,liu2011power,6008569,shin2012adaptive,ramos2011page,6005358,tian2011optimal,seok2011migration,baek2012dual,Kwon2012CSA,6522338,kim2012hybrid,sohail2012migrantstore,rwa2011springer,meza2012enabling,park2011powericnit,yoon2012row,lee2011energy,wu2011improving,hsieh2012double,vamsikrishna2012write,6374793,hwang2013hmmsched}. These techniques aim to achieve the best of both DRAM and PCM, viz. the  short latency and high write endurance of DRAM and low leakage power and high density of PCM.

Some techniques convert PCM write operation to read-before-write (or data comparison write) operation to reduce write energy \cite{wang2011energy,chen2012energy,cho2009flip,yang2007low,mirhoseini2012coding,joshi2011mercury}. These techniques, referred to as ``differential write'' based techniques, read out the old value in the PCM array before writing the new one and compare them to write only those bits that need to change. Some researchers propose task-scheduling based techniques to address the challenges in hybrid DRAM-PCM based main memory \cite{hwang2013hmmsched,tian2011optimal}.

Several other techniques are based on reducing write-traffic to PCM memory (e.g. \cite{xu2009data}). Some researchers propose last level cache management techniques for improving energy efficiency of PCM main memory \cite{barcelo2012energy,ferreira2010increasing}. 
Other researchers have proposed compression based techniques to reduce write-traffic to PCM \cite{baek2012dual,Du2013DCO}. Several other techniques aim to address the write latency issue, and its harmful impact on read latency arising due to bank conflicts and try to utilize  write locality to coalesce all possible changes to the data by using buffers before they are finally written to PCM \cite{zhang2009exploring,lee2009architecting}. Several wear-leveling techniques which work by reducing the number of writes to PCM also generally save write energy. 

PCM also offers the ability to store multiple bits per cell and several researchers propose techniques to achieve this in an energy efficient manner \cite{joshi2011mercury,Li2013CDW,wang2011energy,jiang2012fpb,dong2011adams}. Finally, some researchers have proposed architecture or device-level simulators for studying non-volatile memories \cite{6296505,dong2012nvsim,zhu2013scm,li2012sim,dong2009pcramsim}, which facilitate study of PCM.


\section{Energy Saving Techniques}\label{sec:est}

The read power and delay of PCM are in the same range as that of DRAM, however, its write power is significantly higher, which can be several times that of DRAM  \cite{yang2007low,zhou2009durable}. 
In contrast, for DRAM, both read and write times are equivalent. Writing to a PCM cell  requires high current density over a large period of time. Hence, to ensure correct operation, hard limits on the number of simultaneous writes must be enforced which limits write throughput and overall performance. Thus, failure to save write energy may nullify the energy saving advantage gained due to low leakage power of PCM. 

Further, large power consumption of PCM can have deleterious effect on its operation. It may lead to violating power limits, which may in turn lead to voltage drops in the power supply or excessive currents flowing through the processor. It may increase the temperature which may further increase the leakage energy consumption of other components of the system. It may also create logical errors, incomplete PCM phase transitions, PCM read errors, etc. which may lead to chip failures or chip-aging. Thus, power management of PCM is extremely important. In this section, we review several techniques for managing power consumption of PCM.

Hay et al. \cite{hay2011preventing} propose a technique to reduce write power consumption of PCM banks. Their technique is based on the observation that typically only a small portion of the bits (for example, less than 25\% on average) are written to which consume power. Their technique monitors the number of bits that will change on a write and hence, need to be written. This gives an estimate of the number of bits and hence, amount of power consumed in a write. Then,  to not exceed the power budget, the memory controller issues writes only when there is enough power to support them.

Cho et al. \cite{cho2009flip} propose a technique named, Flip-N-Write to improve PCM write bandwidth, write energy, and write endurance under an instantaneous write power constraint. Their technique works on the observation that many bit-writes to PCM are redundant. Their technique replaces a write operation with read-modify write operation to skip writing a bit if the bit being written is same as the originally stored bit. Further, to restrict the maximum number of bits which are written, they use a ``flip'' bit. If storing the flipped value of data requires less number of bit-write operations, their technique stores the data in flipped form and changes the flip bit to ON.  Using their technique, the write bandwidth is doubled, which also improves the write endurance and reduces the write energy. 

Lee et al. \cite{lee2009architecting} propose using multiple row-buffers inside a PCM chip, which reduces the read latency and also the write energy through write coalescing. Multiple writes to the same location are absorbed in the buffers, thus resulting in much smaller number of  write-backs to the PCM array. They also propose a technique which uses multiple dirty bits in the cache blocks to enable partial writes. Using this, the number of bit updates are reduced by not writing untouched, clean data portion in a dirty cache block to the main memory when the cache block is replaced. Their techniques also increase the lifetime of PCM.

Zhang et al. \cite{zhang2009exploring} study PCM in the context of 3D die-stacking. Using analytical and circuit-level modeling for PCM characterization, they show that  the programming power of PCM cells can be reduced as the chip temperature is elevated. This high-temperature friendly operation of PCM can be advantageously used to design 3D die-stacking memory systems.  They propose a hybrid memory design where a large portion of PCM is used as a primary memory space and a small portion of DRAM is used as a write-buffer to reduce the number of writes to PCM. They also propose an OS-level paging scheme that takes into account the memory reference characteristics of applications and migrates the hot-modified pages from PCM to DRAM so that the life time degradation of PCM is alleviated. Their technique also improves the energy efficiency of the memory system.

 Ferreira et al. \cite{ferreira2010increasing} propose a cache replacement policy for saving PCM main memory energy. Their approach aims to reduce the write-back traffic to main memory. The policy is called $N$-Chance where $N$ can be varied. This policy evicts the least recently accessed clean page from cache, unless all of the $N$ least recently accessed pages are dirty, if so, it evicts the least recently accessed page. For the case when $N =1 $, this policy becomes the conventional LRU (least-recently used) policy. They have shown that for a proper choice of $N$, their policy can be significantly better than the LRU policy.

 Hu et al. \cite{Hu2013WAR} propose a technique for reducing the number of writes to PCM main memory. Their technique is based on data migration and re-computation. In an embedded CMP (chip multiprocessor) having scratch-pad memory, their technique migrates data  to the scratch-pad memory of a different core to avoid write-backs of shared data. Thus, by temporarily storing the data on scratch-pad, their technique reduces the number of write-backs.  Their technique uses program analysis to determine when and where the data should be migrated. They also propose data re-computation to  reduce the number of write activities by discarding the data which should have been written back to the main memory and recomputing these data when they are needed.  They model the problem of data migration as a shortest path problem. Also, they propose an approach to  find the optimal data migration path with minimal cost for both dirty data and clean data.

Qureshi et al. \cite{qureshi2009scalable} propose a hybrid memory design where PCM memory is augmented with a small DRAM that acts as a ``page cache'' for the PCM memory. The page cache buffers frequently accessed pages and thus helps performance and improves PCM endurance by reducing the number of writes to PCM with write combining and coalescing. Further, at cache line level, only the lines modified in a page are written to the main memory. Finally, at block-level, swapping is used for achieving wear-leveling. Their technique also reduces the page faults which improves the performance of the system. However, when the applications have poor locality, the advantage of using page cache reduces.


Liu et al. \cite{liu2011power} study the variable partitioning problem on a hybrid main memory designed with PCM and DRAM for DSP systems \cite{liu2011power}. They propose ILP (integer linear programming) formulations and heuristic algorithms, such that the energy efficiency of PCM can be leveraged while also minimizing the performance and lifetime degradation caused by PCM writes.

Bock et al. \cite{bock2011analyzing} propose a technique to save PCM energy and increase its endurance by avoiding useless write-backs. They define a write-back to a lower level cache to be useless when the data that are written back are not used again by the program. As an example, a useless write-back results when a dirty cache line (block) that belongs to a dead memory region is evicted from the cache. Their technique assumes that suitable schemes can be employed to detect dead memory regions in different parts of memory, such as heap, stack and global memory. Assuming that such information is available, their technique estimates the maximum energy savings that could be achieved by avoiding useless write-backs. Further, since writes are not on critical path of execution, avoiding useless write-backs does not have a significant influence on performance.

Mirhoseini et al.  \cite{mirhoseini2012coding} propose a coding-level technique for saving PCM energy, which is based on the observation that PCM set and reset energy costs are not equal. Their technique aims at minimizing the energy cost of rewriting to PCM by designing low overhead data encoding methods. Their encoding scheme utilizes PCM bitwise manipulation ability during the word overwrites such that only the bits which are changing for the new word compared to the original word would require overwriting. They propose an ILP (integer linear programming) method and employ dynamic programming to produce codes for uniformly distributed data.

Yoon et al. \cite{yoon2012exploring} use a design space exploration approach for finding the optimal configuration of cache hierarchy in a system with non-volatile memory.  Use of non-volatile devices enables designing caches with higher capacity and hence, the cache hierarchy in modern systems is becoming deeper with L4 and L5 caches (e.g. off-chip DRAM caches).  They consider both performance and power while performing the experiments. They consider cache hierarchy of different depth, where the cache at any level can be designed with either SRAM, DRAM or PCM. Also the main memory could be designed with DRAM or PCM. They observe that a large last level cache (LLC) designed with PCM can improve energy efficiency by reducing the costly off-chip accesses. Also,  deep  cache hierarchies are less energy  efficient than flat hierarchies (2 or 3 levels).

Seok et al. \cite{seok2011migration} propose a migration based page caching technique for PCM-DRAM hybrid main memory system. Their technique aims to overcome the problem of the long latency and low endurance of PCM. For this, read-bound access pages are kept in PCM and write-bound access pages are kept in DRAM. Their technique uses separate read and write queues for both PCM and DRAM and uses page monitoring to make migration decisions.
Write-bound pages are migrated from PCM to DRAM and read-bound pages are migrated from DRAM to PCM. The decision to migrate is taken as follows: when a write access is hit and the accessed page is in PCM write queue, it is migrated. Similarly, if a read access is hit and the accessed page is in the DRAM read queue, it is migrated.

Dhiman et al. \cite{dhiman2009pdram} propose a hybrid main memory system composed of DRAM and PCM. Their memory system exposes DRAM and PCM addressability to the software (OS). In their technique, data placement is performed based on the write frequency to data.  If the number of writes to a PCM page exceeds a threshold, the contents of the page are copied to another page (either in DRAM or PCM) for achieving PCM wear-leveling.

Fang et al. \cite{fang2012softpcm} propose a technique called ``SoftPCM'' which utilizes the error tolerance characteristic of video applications to relax the accuracy of write operations. It is well-known that several multimedia applications have the inherent ability of error tolerance \cite{pande2013embedded,sood2006novel,mittal2010content,pande2009network,pande2009dynamic,chan2012temporal}. Thus, a slight error in multimedia data may not be perceived by human end-users. Their technique leverages this fact and provisions that if the stored old data in PCM are very close to the new data to be written,  the write operation is cancelled and the old data are taken as the new data. This leads to significant reduction in write-traffic which also reduces the energy consumption of PCM.

Qureshi et al. \cite{qureshi2012preset} propose a technique to alleviate the problem of slow writes. Their technique works on the observation that PCM  writes are slow only in one direction (SET operation) and are almost as fast as reads in the other direction (RESET operation). Thus, a write operation to a line in which all memory cells have been SET before the write will consume much less time. Based on this, their technique pro-actively
SETs all the bits in a given memory line much before the anticipated write to that memory line. As soon as a line becomes dirty in the cache, their technique initiates a SET request for that line, which allows a large window of time for the SET operation to complete.

As the demand for data increases \cite{mcdonagh2013toward,jana2013mobile,pande2013video}, the size of main memory required will also increase and this would require effective management of main memory. Zhou et al. \cite{zhou2009durable} propose a technique which works by removing redundant bit to reduce the unnecessary bit writes to PCM. Their technique performs a read before write, and writes only those bits to PCM which have changed.


Ramos et al. \cite{ramos2011page} propose a PCM-DRAM hybrid design for improving energy efficiency of main memory. Their technique uses a  hardware-driven page placement policy. Their policy leverages the memory controller to monitor program access patterns and uses this information to migrate pages between DRAM  and PCM, and translate the memory addresses coming from the cores. Further, the operating system periodically updates its page mappings based on the translation information used by the memory controller. Since most frequently accessed pages reside in DRAM, the high write latency of accessing PCM is avoided.

Qureshi et al. \cite{qureshi2010improving} propose a write cancellation and write pausing technique to indirectly improve the PCM read performance. Although a higher value of write latency can be tolerated using buffers and large write bandwidth, once a write request is scheduled for service to a PCM bank, a subsequent read access to the same bank needs to wait until the write access has completed. Thus, the slow write can increase the effective latency of read accesses and since read accesses are latency-critical, this may severely affect the program performance. Write cancellation policy aborts an on-going write if a read request arrives to the same bank and  the write operation is not close to completion. It avoids aborting an ongoing write that has completed more than a threshold percentage of its service time and this threshold can be adapted during runtime. 
 
A limitation of write-cancellation technique is that it requires some writes to be re-executed which incurs power and bandwidth overhead. To avoid this overhead, Qureshi et al. \cite{qureshi2010improving} propose write pausing technique. This technique utilizes fundamental  characteristic  of PCM that most multi-bit PCM devices use iterative write algorithms.  In each iteration data are written and the current state of the device is compared with the desired state.    Write pausing allows iterative write algorithms to potentially pause a write request at the end of each write iteration, complete a pending read request, and then resume the paused write request.

Tian et al. \cite{tian2011optimal} present a task-scheduling based technique for addressing the challenges of hybrid DRAM-PCM main memory. They study the problem of task-scheduling, assuming that a task should be entirely placed in either PCM bank or DRAM bank. Their approach works for different optimization objectives such as 1. minimizing the energy consumption of hybrid memory for a given PCM and DRAM size and given PCM endurance 2. minimizing the number of writes to PCM for a given PCM and DRAM size and given threshold on energy consumption and 3. minimizing PCM size for a given DRAM size, given threshold on energy consumption and PCM endurance.

In context of PCM-DRAM hybrid main memory, Meza et al. \cite{meza2012enabling} propose a technique for efficiently managing the metadata (such as tag, LRU, valid, and dirty bits) for data in a DRAM cache at a fine granularity. Their technique uses the observation that storing metadata off-chip in the same row as their data can exploit DRAM row buffer locality; also it reduces the access latency from two row buffer conflicts (one for the metadata and another for the datum itself). Based on this, their technique only caches the  metadata for recently accessed rows on-chip using a small buffer. Since metadata needed for data with temporal or spatial locality is cached on-chip, it can be accessed with   the same latency as an SRAM tag store. This provides better energy efficiency than using a large SRAM tag store.

Yoon et al. \cite{yoon2012row} propose  row-buffer locality aware caching policies for hybrid PCM-DRAM main memories. Their technique works on the observation that both DRAM and PCM have row buffers, with (nearly) same latency and bandwidth. However, the cost of row buffer misses in terms of latency, bandwidth, and energy is much higher in PCM  than in DRAM. Based on this, their technique avoids allocating in PCM data that frequently causes row buffer misses. Such data are allocated (cached) in DRAM, whereas the data that frequently hits in the row-buffer are stored in PCM. Further, since PCM has much higher write latency/power than read latency/power, their technique uses a caching policy such that the pages that are written frequently are more likely to stay in DRAM.

Huang et al. \cite{huang2011register} propose a register-allocation based technique with re-computation to reduce the number of store instructions to non-volatile memory. Register allocation refers to multiplexing a large number of target program variables onto a small number of physical registers. The less the number of physical registers a processor contains, the more number of spills will be generated. Each spill is mapped to one store instruction and one (or few) load instructions during the compilation process. Traditional register
allocation process does not distinguish read and write activities and does not try to minimize writes. Huang et al. use graph-coloring approach to extend traditional register allocation technique with re-computation to reduce non-volatile memory write activities by reducing store instructions. Their technique discards a set of carefully-selected actual spills and re-computes them when they are needed

 Although PCM MLC devices offer more density than SLC (single level cell) devices, they also present significant challenges. For MLC devices to work properly, precise reading of reistance values is required. As the number of levels increase, the resistance region assigned to each data value decreases significantly. Thus, the read latency of MLC devices may increase linearly or exponentially with the number of bits. Qureshi et al. \cite{qureshi2010morphable} present a memory architecture which aims to achieve the latency, lifetime and energy of SLC devices in the common case, while still achieving the high memory capacity of MLC device. Their technique divides the main memory in two regions, one with high-density, high-latency which uses MLC mode, and another with low-latency, low-density that uses half the number of bits per cell than high-density region. By tracking the memory requirements of the workloads, their technique adapts the fraction of both regions. When the workload requires high memory capacity, the system uses capacity benefits of MLC device. When the workload requirements can be satisfied with SLC (or fewer bits per level cell), the system increases the size of SLC region to avoid increased energy and latency. This is achieved by restricting the  number of levels used in a MLC device to emulate a fewer bits per cell device. To avoid high latency for frequently accessed pages, the system transfers a page from high-density region to low-density (faster) region when the page is accessed.

  Lee et al. \cite{6005358} propose a memory management technique for  hybrid PCM-DRAM memory to hide the slow write performance of PCM. Their technique uses methods such as dirty bit clearing and frequency accumulation to accurately  estimate future write references. They observe that  using write history alone performs better than using both read and write history in estimating future write references.  Also, by using temporal locality and frequency characteristics, more accurate estimates of write references can be obtained. Based on these observations, they propose a  page replacement algorithm called CLOCK-DWF (CLOCK with dirty bits and write frequency) that reduces the number of PCM write operations by using DRAM to absorb most of the write references.

  Wang et al. \cite{wang2011energy} propose a technique for mitigating the energy overhead of MLC PCM devices. Their technique works on the observation that there are significant value-dependent
energy variations in programming MLC PCM. Thus, by using data encoding, the write energy can be reduced. In a 2-bit PCM, there are four states, viz. 00, 01, 10 and 11. They show that programming states 00 and 11 require significantly less energy than programming the other states. Thus, data encoding is used to increase the 00 and 11 states in writing the data. They also use data comparison write (DCW) approach to enhance the effectiveness of the data encoding scheme.

  Joshi et al. \cite{joshi2011mercury} present a circuit and microarchitecture-level technique to address the high write latency of MLC-PCM.  The write latency and energy of PCM vary significantly with target resistance level and the initial state of PCM cell. Their technique adapts the programming scheme of MLC PCM by taking into consideration the initial state of the cell, the target resistance to be programmed and the effect of process variation on the programming current profile of the MLC.  For states mapped at lower resistance values, they use single reset pulse programming and for states mapped at higher resistance values, they use staircase programming. Also, data comparison writes (DCW) is used to enhance the effect of their technique. Also, when the cell is already present in the stable completely set state, their technique skips initialization sequence for programming, which further improves the write latency and energy saving.
  
  Xu et al. \cite{xu2009data} propose data manipulation techniques to reduce the write energy of PCM main memory. Their technique works on the observation that PCM read incurs much less energy than PCM writes. Also, write of different value to a PCM cell incurs significantly different energy.   
  Their technique uses selective-XOR operations to bias the data value distribution. For a given word to be written and originally stored word, their technique finds an optimal bit-pattern such that writing a XOR-masked value of word to be written with bit pattern leads to minimum write energy.

\section{Conclusion}\label{sec:conclusion}

Driven by the quest for exascale performance, modern processors, data-centers and supercomputers use large sized memory and limitations of conventional devices forces the designers to explore new avenues of design cache and memory hierarchy. The use of phase change memory (PCM) as a universal choice for main memory  opens us several new challenges. It is clear that while PCM provides a promising alternative to conventional DRAM, it cannot completely replace DRAM due to its limitations. Thus, the computer architects and researchers need to design effective techniques at system, architecture and device level to address the shortcomings of PCM and utilize its strengths.

In this report, we have presented a review of  techniques proposed for power management of phase change memory. We believe that our work will be useful for both beginners and experts in the field of PCM. Also, it will help them in gaining insights into the working of architectural techniques of PCM power management and encourage them to improve these techniques even further.



\ifCLASSOPTIONcaptionsoff
  \newpage
\fi



\bibliographystyle{IEEEtran}
\bibliography{PhDReferences}

\end{document}